\newif\ifcikmdraft
\ifdefined\cikmdraft
  \cikmdrafttrue
\else
  \cikmdraftfalse
\fi

\ifcikmdraft
  \documentclass[sigconf,review]{acmart}
\else
  \documentclass[sigconf]{acmart}
\fi
\usepackage{enumitem}
\usepackage[ruled,vlined, linesnumbered]{algorithm2e}
\usepackage{booktabs}
\usepackage{pifont}
\usepackage{multirow}
\usepackage{threeparttable}
\usepackage{booktabs}
\usepackage{threeparttable}
\usepackage{tabularx}
\usepackage{array}
\usepackage{ragged2e}

\newcolumntype{Y}{>{\RaggedRight\arraybackslash}X}

\newcommand{\revise}[1]{{\color{red}  #1}}
\newcommand{\cmark}{\ding{51}}   
\newcommand{\xmark}{\ding{55}}   
\newcommand{\paratitle}[1]

\usepackage{booktabs}
\usepackage{flafter}
\usepackage{multirow}
\usepackage{microtype}
\usepackage{colortbl}
\usepackage{xspace}
\usepackage{enumitem}
\usepackage{tcolorbox}

\newtcolorbox{findingboxenv}{
  colback=black!6,
  colframe=black!18,
  boxrule=2pt,
  arc=2mm,
  left=5pt,
  right=5pt,
  top=5pt,
  bottom=5pt,
  before skip=6pt,
  after skip=8pt
}

\copyrightyear{2027}
\acmYear{2027}
\setcopyright{cc}
\setcctype{by}
\acmConference[KDD 2027] {Proceedings of the 33rd ACM SIGKDD Conference on Knowledge Discovery and Data Mining V.1}{August 1--5, 2027}{San Jose, CA, USA.}
\acmBooktitle{Proceedings of the 33rd ACM SIGKDD Conference on Knowledge Discovery and Data Mining V.1 (KDD 2027), August 1--5, 2027, San Jose, CA, USA}
\acmISBN{979-8-4007-3134-1/2027/08}
\acmDOI{10.1145/3862868.3871538}

\begin{document}

\title{KuaiLive-M3: A Multi-Modal, Multi-Domain, and Multi-Feedback Dataset for Live Streaming Recommendation}

\author{Ke Guo}
\author{Changle Qu}
\affiliation{
  \institution{\mbox{Gaoling School of Artificial Intelligence}\\Renmin University of China}
  \city{Beijing}
  \country{China}
  }
\email{\{imgkkk2004,changlequ\}@ruc.edu.cn}

\author{Jiayaqi Cheng}
\author{Xiao Zhang}
\authornote{Corresponding author: Xiao Zhang (e-mail: zhangx89@ruc.edu.cn).}
\affiliation{
  \institution{\mbox{Gaoling School of Artificial Intelligence}\\Renmin University of China}
  \city{Beijing}
  \country{China}
  }
\email{\{jiayaqicheng, zhangx89\}@ruc.edu.cn}

\author{Shijun Wang}
\author{Xiaoyu Zhang}
\affiliation{
  \institution{Kuaishou Technology}
  \city{Beijing}
  \country{China}
  }
\email{\{wangshijun03,zhangxiaoyu\}@kuaishou.com}

\author{Xueliang Wang}
\author{Le Zhang}
\affiliation{
  \institution{Kuaishou Technology}
  \city{Beijing}
  \country{China}
  }
\email{\{wangxueliang03,zhangle07\}@kuaishou.com}

\author{Lantao Hu}
\affiliation{
  \institution{Kuaishou Technology}
  \city{Beijing}
  \country{China}
  }
\email{hulantao@kuaishou.com}

\author{Jun Xu}
\affiliation{
  \institution{\mbox{Gaoling School of Artificial Intelligence}\\Renmin University of China}
  \city{Beijing}
  \country{China}
  }
\email{junxu@ruc.edu.cn}

\renewcommand{\shortauthors}{Ke Guo et al.}

\begin{abstract}

Live streaming has become a major form of online content consumption, driving increasing interest in live streaming recommendation.
Unlike conventional item recommendation, live streaming recommendation must model continuously evolving content and real-time user interactions.
However, existing public datasets suffer from three major limitations: they provide limited access to temporally evolving multimodal live content, overlook users' cross-domain interactions between short videos and live streams, and contain only implicit behavioral signals without explicit feedback that captures users' perceived content quality and satisfaction. 
These limitations prevent existing benchmarks from faithfully reflecting real-world live streaming scenarios and hinder comprehensive research on live streaming recommendation.
To address these limitations, we introduce \textbf{KuaiLive-M3}, a \textbf{multi-modal}, \textbf{multi-domain}, and \textbf{multi-feedback} dataset for live streaming recommendation, collected from Kuaishou, a leading live streaming and short video platform in China. KuaiLive-M3 covers 21,938 users and contains 35 million live streaming interactions and 111 million short video interactions, with fine-grained timestamps and diverse user behaviors. 
It further provides approximately 88 million timestamped segment-level multi-modal embeddings that capture the temporal evolution of live streaming content, as well as 25,403 questionnaire-based feedback records that bridge implicit user behaviors and explicit user preferences.
Based on these unique signals, we establish benchmarks for cross-domain recommendation, live stream highlight prediction, and questionnaire-enhanced recommendation. 
Extensive experiments with representative baselines demonstrate that KuaiLive-M3 provides a challenging and realistic benchmark for future live streaming recommendation research. The results further highlight the importance of modeling temporally evolving content, transferring user preferences across domains, and bridging the gap between implicit behaviors and explicit user feedback. The dataset and benchmark code are publicly available at~\textcolor{blue}{\url{https://imgkkk574.github.io/KuaiLive-M3/}}.

\end{abstract}



\maketitle

\section{Introduction}

\begin{figure}
    \centering
    \includegraphics[width=\linewidth]{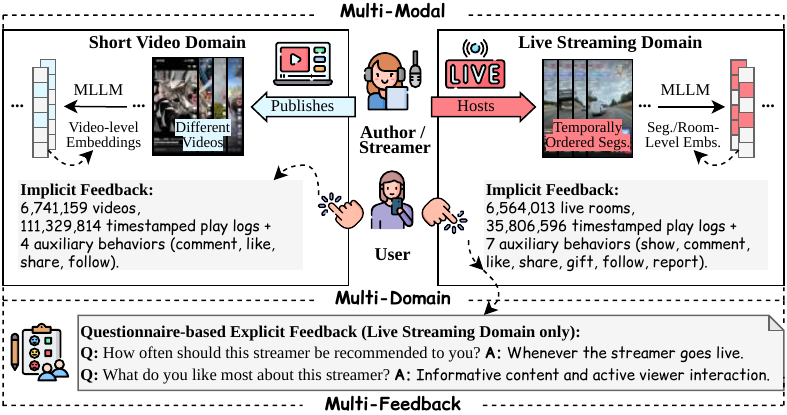}
    \caption{Overview of the KuaiLive-M3 Dataset.}
    \vspace{-0.2cm}
    \label{fig:intro}
    \vspace{-0.5cm}
\end{figure}

Live streaming has become an increasingly prominent form of online content consumption~\cite{wang2026onelive,zhang2026bridging}. Its real-time interactivity, dynamically evolving content, and diverse user behaviors pose distinctive recommendation challenges that are rarely encountered in conventional recommendation settings~\cite{qu2025kuailive,guo2026room,liu2025llm,qu2025bridging,shi2026ssrlive,zhao2025towards}. 
In a live streaming platform, streamers continuously broadcast content in live rooms, while users navigate among active rooms and interact through watching, commenting, liking, and gifting. The content of a live room evolves over time and involves multiple modalities, such as real-time video and audio, making temporally aligned multi-modal signals essential for assessing its relevance to users at different moments~\cite{deng2024multimodal-KLive,liang2024ensure,deng2023contentctr}. Moreover, live streaming services are often deeply integrated with short video ecosystems, as exemplified by Kuaishou\footnote{\url{https://www.kuaishou.com}}. 
On Kuaishou, 99.44\% of live streaming viewers also watch short videos, 98.07\% of streamers also publish short videos.
These substantial behavioral and content overlaps make short-video--live-streaming cross-domain recommendation~(CDR) an important problem with considerable practical value~\cite{qu2025bridging,zhang2026bridging}.

Despite its practical importance, research on live streaming recommendation remains constrained by the lack of public, real-world datasets that jointly capture temporally evolving multi-modal content, cross-domain behaviors, and explicit user feedback. Existing public datasets primarily focus on the live streaming domain alone and rely predominantly on implicit interaction signals~\cite{rappaz2021recommendation-liverec,yu2021leveraging-LSEC,qu2025kuailive}. In particular, KuaiLive~\cite{qu2025kuailive} advances this line of research by providing realistic, timestamped sequences of user interactions in an interactive live streaming environment. However, it does not capture users' behaviors in the closely related short video domain, provide segment-level multi-modal representations of evolving live content, or include questionnaire-based feedback reflecting users' explicit judgments. Consequently, existing datasets cannot adequately support multi-modal, cross-domain, and human-centered research for live streaming recommendation.

To bridge this gap, we introduce \textbf{KuaiLive-M3}, a large-scale \textbf{multi-modal}, \textbf{multi-domain}, and \textbf{multi-feedback} dataset for live streaming recommendation, collected from Kuaishou. As one of the largest live streaming and short video platforms in China, Kuaishou provides a natural environment for studying the interplay between these two closely related domains.
Figure~\ref{fig:intro} provides an overview of KuaiLive-M3, highlighting its multi-domain interactions, multi-granularity multi-modal representations, and implicit and questionnaire-based explicit feedback.
Compared with existing public live streaming datasets, KuaiLive-M3 offers the following advantages: 
\textbf{(1) Real-world cross-domain behaviors.}
KuaiLive-M3 records the activities of shared users and streamers across the live streaming and short video domains, enabling realistic cross-domain recommendation and preference-transfer research.
\textbf{(2) Temporally evolving multimodal content.}
It provides timestamped segment-level and room-level multimodal embeddings for live streams, together with video-level embeddings for short videos. These representations support content-aware recommendation and the modeling of temporal content evolution within live rooms.
\textbf{(3) Questionnaire-based explicit feedback.}
Beyond implicit behavioral signals, KuaiLive-M3 includes real-world questionnaire responses that reflect users' explicitly expressed preferences and perceived recommendation quality, enabling the study of potential discrepancies between implicit engagement and explicit feedback.
\textbf{(4) Fine-grained and realistic recommendation contexts.}
The dataset contains detailed impression and interaction logs, comment text, live room lifecycle timestamps, rich room- and video-level metadata, and diverse behavior types. These signals enable researchers to reconstruct time-varying candidate sets and establish realistic recommendation and evaluation settings.

Building upon these unique characteristics, we establish a systematic benchmark on KuaiLive-M3 covering three representative tasks. 
First, we study cross-domain recommendation to evaluate whether short video behaviors can improve live streaming recommendation. 
Second, we introduce live stream highlight prediction, which leverages timestamped segment-level embeddings to identify engaging moments from temporally evolving live content. 
Third, we conduct questionnaire-enhanced recommendation to investigate how explicit user feedback can complement implicit behavioral signals. 
For each task, we provide standardized data preprocessing, evaluation protocols, and representative baselines. By enabling reproducible research on cross-domain preference transfer, temporal multimodal modeling, and implicit--explicit feedback alignment, KuaiLive-M3 bridges important gaps between public benchmarks and real-world live streaming recommendation systems.
Beyond the benchmarked tasks, KuaiLive-M3 further supports research on staytime prediction~\cite{zhao2025towards}, generative recommendation~\cite{wang2026onelive, shi2026ssrlive}, and other emerging directions in the live streaming domain. 
We believe KuaiLive-M3 will serve as a valuable resource for advancing live streaming recommendation research.


\begin{table*}[t]
\centering
\caption{Comparison between KuaiLive-M3 and existing public live streaming recommendation datasets. \cmark~and~\xmark~denote full support and not available, respectively. }
\label{tab:dataset_comparison}
\begin{threeparttable}
\small
\begin{tabular}{llccccc}
\toprule
\textbf{Category} & \textbf{Property}
  & \textbf{LiveRec$^{1}$}
  & \textbf{LSEC$^{2}$}
  & \textbf{KLive}
  & \textbf{KuaiLive}
  & \textbf{KuaiLive-M3} \\
\midrule
\multirow{2}{*}{Multi-Domain}
  & Live Streaming         & \cmark  & \cmark & \cmark & \cmark & \cmark \\
  & Short Video            & \xmark & \xmark & \xmark & \xmark & \cmark \\
\midrule
\multirow{3}{*}{Multi-Modal}
  & Room-Level      & \xmark & \xmark & \xmark & \xmark & \cmark \\
  & Segment-Level   & \xmark & \xmark & \cmark & \xmark & \cmark \\
  & Short Video     & \xmark & \xmark & \xmark & \xmark & \cmark \\
\midrule
\multirow{3}{*}{Multi-Feedback}
  & Questionnaire-Based      & \xmark & \xmark & \xmark & \xmark & \cmark \\
  & Negative                 & \xmark & \xmark & \xmark & 12.7M & 92.5M \\
  & Multi-Behavior   & \xmark & \xmark & \xmark & \cmark &\cmark \\
\midrule
\multirow{5}{*}{General}
  & \#Users                   & 15.5M    &  202.9k  & \xmark  & 23.8k  & 21.9K \\
  & \#Streamers (Authors)               &  465k    &   7.4k   &  9.9k   & 452.6k & 2.2M \\
  & \#Rooms                   &  -$^3$       &   \xmark &  17.8k  & 11.6M  & 6.6M \\
  & \#Interactions            &  124M    &  5.4M    & \xmark  & 5.4M   & 35.8M \\
  & Lifecycle Timestamp       &  \xmark  & \xmark   &  \xmark & \cmark & \cmark \\
\bottomrule
\end{tabular}
\begin{tablenotes}
\footnotesize
    \item[1] LiveRec has two versions: Bench. and Full. The reported statistics are based on the Full version.
    \item[2] LSEC has two versions: Small and Large. The reported statistics are based on the Large version.
    \item[3] LiveRec does not report the number of live rooms; it only provides segment-level interaction records within live rooms.
\end{tablenotes}
\end{threeparttable}
\end{table*}

\section{Related Work}
\paragraph{\textbf{Live Streaming Recommendation.}}
With the growing popularity of live streaming, recommendation research in this domain has attracted increasing scholarly attention~\cite{shi2026ssrlive, guo2026room, deng2023contentctr, liang2024ensure, rappaz2021recommendation-liverec, lu2025liveforesighter}. Given the tight integration between live streaming and short video platforms, leveraging short video interaction logs to enhance live streaming recommendation via cross-domain transfer has emerged as an important research topic~\cite{qu2025bridging, zhang2026bridging, li2026farm}. 
Meanwhile, the inherently multi-modal content of live streams has motivated methods that jointly model visual, textual, and acoustic signals for better content and preference representations~\cite{deng2023contentctr,deng2024multimodal-KLive, deng2024mmbee,zhang2026bridging,liu2025llm,xi2023multimodal}.
A closely related problem is live stream highlight prediction, which aims to identify engaging segments from temporally evolving live content~\cite{deng2024multimodal-KLive,zhao2022antpivot}.
While these efforts have advanced the field, existing works are mostly benchmarked on proprietary industrial data, and the lack of publicly available datasets that jointly provide multimodal content and cross-domain interaction data continues to hinder reproducible research in this area. 

\paragraph{\textbf{Live Streaming Datasets.}}
In contrast to the growing methodological interest, publicly accessible datasets for live streaming research remain scarce.
The Twitch dataset~\cite{rappaz2021recommendation-liverec} records time-varying user--streamer consumption but provides limited room-level content and interaction information.
LSEC~\cite{yu2021leveraging-LSEC} models interactions among users, streamers, and products in live streaming e-commerce, with an emphasis on product recommendation and purchasing behavior.
For live stream highlight detection, KLive~\cite{deng2024multimodal-KLive} offers segment-level multi-modal features and dense audience-derived highlight scores, but does not provide the comprehensive interaction logs required for recommendation.
More recently, KuaiLive~\cite{qu2025kuailive} provides real-world, timestamped interactions among users, streamers, and live rooms with precise room lifecycle information, yet remains limited to the single live streaming domain without temporally aligned multi-modal content or questionnaire-based explicit feedback.
The comparison is presented in Table~\ref{tab:dataset_comparison}.
By jointly providing interaction logs and multi-modal representations across both domains, together with questionnaire-based feedback, KuaiLive-M3 enables more comprehensive research into the live streaming ecosystem.

\begin{table}[t]
    \centering
    \caption{Questionnaire design for explicit feedback collection.}
    \label{tab:questionnaire_design}
    \footnotesize
    \begin{tabular}{c m{3cm} m{3cm}}
    \toprule
    Choice & Q1: How often should this streamer be recommended to you? & Q2: Follow-up (Open-ended) \\
    \midrule
    A & Recommend whenever the streamer goes live.
    & What do you like most about this streamer? \\
    \midrule
    B & Recommend occasionally.
    & In what aspects could this streamer improve? \\
    \midrule
    C & Do not recommend this streamer.
    & Why do you no longer wish to see this streamer? \\
    \bottomrule
    \end{tabular}
    \vspace{-0.5cm}
\end{table}

\begin{table*}[h!]
    \centering
    \caption{Statistics of the proposed KuaiLive-M3. Symbols \ding{172}, \ding{173}, \ding{174}, \ding{175}, \ding{176}, \ding{177}, \ding{178} represent seven types of actions: show, comment, like, share, gift, follow, and report. \#R, \#V, and \#S denote the number of rooms, videos, and live segments, respectively.}
    \label{tab:stat}
    \small
    \begin{tabular}{lccccccc}
    \toprule
    Domain & \#Users & \#Authors & \#Items & \#Play & \#Questionnaire & Multi-modal Emb. & Action types\\
    \midrule
    Live & \multirow{2}{*}{21,938} & \multirow{2}{*}{2,185,201} & 6,564,013 (\#R)  & 35,806,596 & 25,403 & 6,502,107 (R), 88,223,794 (\#S) &  \ding{172}, \ding{173}, \ding{174}, \ding{175}, \ding{176}, \ding{177}, \ding{178} \\
    Video    &                    &                    & 6,741,159 (\#V) & 111,329,814 & - & 5,498,631 (V) & \ding{173}, \ding{174}, \ding{175}, \ding{177}\\
    \bottomrule
    \end{tabular}
\end{table*}


\section{The KuaiLive-M3 Dataset}
This section provides an overview of KuaiLive-M3. We first describe its data construction process and licensing terms, and then present its key statistics in Section~\ref{sec:stat}. 

\subsection{Data Construction}
We construct KuaiLive-M3 through the following steps:
\subsubsection{User Sampling. }
To ensure that the dataset reflects real-world platform usage, we randomly sample approximately 22,000 active users from Kuaishou. Each sampled user generated at least one play log per week during a four-week period from March 4 to March 31, 2026. No additional sampling constraints are imposed.

\subsubsection{Interaction Collection.} 
We collect fine-grained interactions of the sampled users in both the live streaming and short video domains. 
In the live streaming domain, we retain detailed valid play logs and records of seven auxiliary behavior types, namely show, comment, like, share, gift, follow, and report. Here, show denotes an impression without a subsequent click and can therefore serve as a negative signal.
In the short video domain, we similarly retain play logs and four auxiliary behavior types, including comment, like, share, and follow. All interaction records in both domains are associated with precise timestamps. To maintain a manageable dataset scale, we randomly sample interactions from the short video domain. The resulting dataset encompasses 21,938 users, 2,185,201 authors, 6,564,013 live rooms, and 6,741,159 videos.

\subsubsection{Multi-modal Embedding Extraction.}
We extract multi-modal embeddings using an industrial multi-modal large language model (MLLM) based on MiniCPM-V-2B~\cite{yao2024minicpm} and fine-tuned on domain-specific knowledge from both the live streaming and short video domains.
The inputs to the MLLM include uniformly segmented clips, automatic speech recognition (ASR) transcripts, titles and descriptions of short videos and live streams, and basic author profile information.
For live streaming content, we provide embeddings at two levels of granularity, namely the segment level and the room level. 
Segment-level embeddings are derived from uniformly segmented clips within each live room. 
For scalability, we retain only segments associated with at least one user interaction and reduce their embeddings to 128 dimensions using PCA. 
Room-level embeddings are obtained by aggregating the segment-level embeddings of all retained segments within a live room and are further reduced to 64 dimensions using PCA. 
For short videos, we provide 128-dimensional video-level embeddings.

\begin{figure*}[t]
    \centering
    \includegraphics[width=1\linewidth]{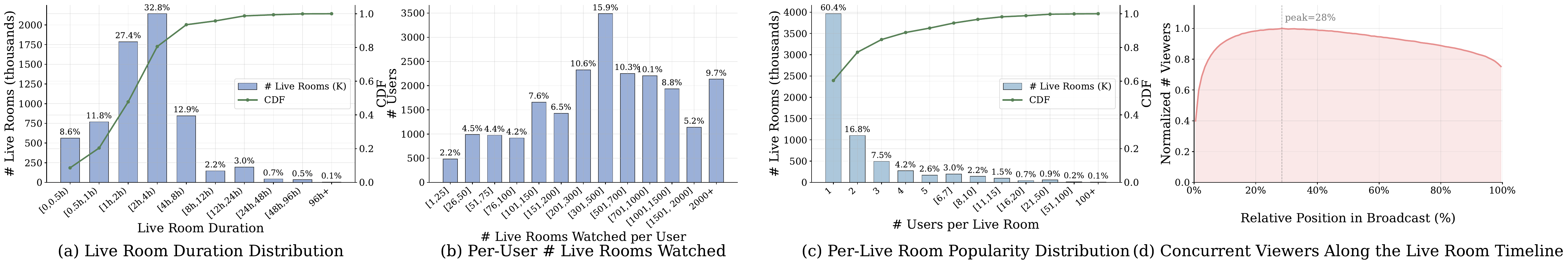}
    \caption{Analysis of interaction patterns in the live streaming domain. 
    (a) shows the distribution of live room durations. 
    (b) illustrates the distribution of distinct live rooms watched per user, with a median of 415. 
    (c) shows the per-room audience size distribution. 
    (d) illustrates the average viewer arrival distribution over the normalized stream lifecycle $\tau \in [0, 1]$.}
    \label{fig:placeholder}
\end{figure*}

\subsubsection{Questionnaire Collection}
To collect explicit feedback, we administer a streamer recommendation frequency preference questionnaire to users in the live streaming domain, distributed uniformly at random among eligible users. 
Each questionnaire consists of two questions: a multiple-choice question asking users to indicate 
their preferred recommendation frequency for the current streamer, followed by an open-ended question soliciting the rationale behind their choice. 
Detailed questionnaire design is presented in Table~\ref{tab:questionnaire_design}, with an illustrative example shown in Figure~\ref{fig:intro}. 
In total, 25,403 questionnaire responses are collected.

\subsubsection{Anonymization and Ethics Statement.}
To protect user privacy and comply with data-release policies, we apply strict anonymization to KuaiLive-M3. 
Specifically, all identifier fields, including user IDs, author IDs, live room IDs, video IDs, and segment IDs, are mapped to anonymized integer identifiers to prevent direct traceability.
All data fields, including questionnaire responses and comment text, have undergone internal ethics review prior to release. The questionnaire was administered to users who were informed of the data collection purpose, and participation was voluntary. No personally identifiable information is retained in the released dataset. This process mitigates privacy risks while preserving the dataset's utility for reproducible research.


\subsubsection{KuaiLive-M3 License.}
Our KuaiLive-M3 is released for non-commercial research and educational use under the Creative Commons Attribution-NonCommercial-ShareAlike 4.0 International License (CC BY-NC-SA 4.0)\footnote{\url{https://creativecommons.org/licenses/by-nc-sa/4.0/}}. 

\subsection{Statistics}
\label{sec:stat}

Table~\ref{tab:stat} summarizes the key statistics of KuaiLive-M3. 
The dataset contains 21,938 users and 2,185,201 authors. 
In the live streaming domain, it includes 35,806,596 play log records associated with 6,564,013 live rooms, together with records of seven  auxiliary behavior types. 
In the short video domain, it includes 111,329,814 play log records associated with 6,741,159 videos, together with records of four auxiliary behavior types.
For multi-modal content representations, the live streaming domain provides 6,502,107 room-level embeddings and 88,223,794 segment-level embeddings, while the short video domain provides 5,498,631 video-level embeddings. In addition, the dataset includes 25,403 questionnaire responses collected from users in the live streaming domain as explicit feedback. 
Detailed dataset schema and field descriptions are provided 
in Appendix~\ref{app:dataset_schema}.

\section{Dataset Analysis}
In this section, we analyze KuaiLive-M3 along two dimensions: interaction patterns in the live streaming domain and cross-domain transfer characteristics between the live streaming and short video domains. These analyses reveal the distinctive behavioral and structural properties of the dataset, offering practical insights for downstream live streaming recommendation tasks.

\subsection{Interaction Patterns}
In this section, we first analyze the interaction patterns in the live streaming domain from the following three aspects:

\subsubsection{Live Room Duration Distribution.}
We first analyze the duration distribution of live rooms. As shown in Figure~\ref{fig:placeholder}~(a), the majority of live rooms (80.6\%) have durations within 4~hours, indicating that most broadcasts are time-bounded sessions. 
Nevertheless, a non-trivial proportion of live rooms exceed 8~hours, and a small number of extreme cases last for more than 24~hours. These exceptionally long sessions typically correspond to companion streams or always-on broadcasts, such as news channels. 
This heavy-tailed duration distribution suggests that temporal modeling is essential for live streaming recommendation, as engagement patterns may vary substantially across streams of different durations.

\subsubsection{User and Live Room Engagement Distribution.}
We next examine the engagement distributions of users and live rooms. 
As shown in Figure~\ref{fig:placeholder}~(b), user engagement follows a broad unimodal distribution that peaks in the $[301, 500]$ interval, with a median of 415 distinct live rooms watched per user during the 4-week observation period.
In contrast, live room popularity exhibits a pronounced long-tail distribution. As shown in Figure~\ref{fig:placeholder}~(c), 60.4\% of live rooms are watched by only one user, whereas only 0.1\% are watched by more than 100 unique users. 
This stark asymmetry highlights the pervasive cold-start challenge faced by most live rooms and underscores the need to develop effective exposure strategies for emerging streamers.

\subsubsection{Temporal Viewer Distribution.}
Figure~\ref{fig:placeholder}~(d) illustrates how viewer attention evolves over the course of a live room. Since live rooms vary widely in duration, we normalize each room to a relative timeline $\tau \in [0, 1]$. We then apply sweep-line aggregation to compute the number of concurrent viewers at each relative position and average the results across all live rooms.
The resulting curve reveals a pronounced front-loaded pattern, where viewer concentration peaks sharply during the early phase of a broadcast and decays monotonically thereafter. 
This temporal asymmetry highlights the importance of timely recommendation delivery. 
A system that fails to surface a live room during its early high-engagement window may miss the primary opportunity for audience acquisition. 
Consequently, highlight prediction, which aims to identify promising stages in a room's lifecycle for recommendation to candidate users, emerges as a critical task for live streaming recommendation.

\subsection{Cross-Domain Transfer Properties}
KuaiLive-M3 captures user behaviors in the live streaming and short video domains within a unified 4-week observation window, providing a natural testbed for cross-domain recommendation research. We analyze KuaiLive-M3's cross-domain transfer properties from the following two perspectives:

\subsubsection{Interaction Density Across Domains.} 
KuaiLive-M3 contains 111,329,814 event-level play logs in the short video domain, compared with 35,806,596 play records in the live streaming domain, yielding a ratio of approximately 3:1. 
This difference reflects a platform-level behavioral pattern in which Kuaishou users are substantially more active in the short video domain and consume a larger volume of content per session. 
The richer behavioral signals from the short video domain therefore provide a valuable auxiliary source for alleviating the data sparsity inherent in live streaming. 

\subsubsection{Authors as Cross-Domain Bridges.}
On Kuaishou, authors frequently participate in both domains by publishing short videos and hosting live streams, rather than being confined to a single domain. 
This setting differs from conventional CDR benchmarks, such as Douban~\cite{DTCDR}, in which the two domains involve disjoint item sets and author populations with no shared identities. In such benchmarks, user overlap is typically the only available cross-domain signal.
A distinctive structural property of KuaiLive-M3 is that authors serve as direct bridges between the two domains. 
The dataset contains 2,185,201 unique authors, including 1,061,343 active streamers in the live streaming domain. Among these streamers, 186,895, accounting for 17.6\%, also published short videos during the same observation window. This substantial author overlap enables cross-domain models to leverage shared author identities as alignment signals, supplementing the user co-occurrence signals commonly used in prior work.

Beyond this explicit entity overlap, cross-domain relatedness may also arise between short videos and live rooms without a shared author~\cite{qu2025bridging}. 
For example, a short video may contain clips extracted from a streamer's live broadcast or cover topics closely related to a particular streamer. 
Such cases establish implicit semantic connections without direct authorship links. This setting differs from standard CDR assumptions, in which cross-domain transfer typically relies on overlapping users or items. Taken together, the explicit author overlap and implicit semantic connections motivate the modeling of complex heterogeneous relations among users, authors, videos, and live rooms.

\section{Benchmarked Experiments}

\begin{table*}[htbp]
    \centering
    \caption{Overall performance comparison of single-domain and cross-domain recommendation methods on KuaiLive-M3. The best and second-best results are highlighted in \textbf{bold} and \underline{underlined} fonts, respectively.}
    \label{tab:cross-domain}
    \footnotesize
    \resizebox{\textwidth}{!}{%
    \begin{tabular}{lcccccccccccccc}
    \toprule
    \multirow{2}{*}{\textbf{Metrics}} 
    & \multicolumn{4}{c}{\textbf{Single-domain Methods}} 
    & \multicolumn{8}{c}{\textbf{Cross-domain Methods}}  \\
    \cmidrule(lr){2-5} \cmidrule(lr){6-14}
    & BPRMF & NeuMF & NGCF & LightGCN & CMF & CLFM & EMCDR & CoNet & DCDCSR & DeepAPF & DTCDR & BiTGCF &  MGCCDR \\
    \midrule
    R@10 & 0.0376 & 0.0280 & 0.0344 & \underline{0.0556} & 0.0322 & 0.0172 & 0.0198 & 0.0264 & 0.0320 & 0.0293 & 0.0275 & 0.0379 &   \textbf{0.0817} \\
    N@10 & 0.0965 & 0.0689 & 0.0821 & \underline{0.1358} & 0.0860 & 0.0483 & 0.0537 & 0.0653 & 0.0740 & 0.0694 & 0.0733 & 0.0898 &   \textbf{0.1795} \\
    R@20 & 0.0606 & 0.0435 & 0.0548 & \underline{0.0873} & 0.0511 & 0.0216 & 0.0292 & 0.0415 & 0.0522 & 0.0450 & 0.0431 & 0.0592 &   \textbf{0.1164} \\
    N@20 & 0.0917 & 0.0653 & 0.0783 & \underline{0.1278} & 0.0805 & 0.0398 & 0.0474 & 0.0614 & 0.0715 & 0.0654 & 0.0679 & 0.0852 &   \textbf{0.1651} \\
    R@40 & 0.0934 & 0.0671 & 0.0852 & \underline{0.1313} & 0.0794 & 0.0340 & 0.0422 & 0.0634 & 0.0820 & 0.0683 & 0.0685 & 0.0895 &   \textbf{0.1579} \\
    N@40 & 0.0953 & 0.0683 & 0.0828 & \underline{0.1315} & 0.0829 & 0.0401 & 0.0461 & 0.0639 & 0.0767 & 0.0689 & 0.0711 & 0.0889 &   \textbf{0.1633} \\
    \bottomrule
    \end{tabular}%
    }
\end{table*}

KuaiLive-M3 supports a diverse range of research tasks in live streaming recommendation by providing cross-domain interaction data, multi-modal content embeddings, and explicit questionnaire feedback. 
To demonstrate its research utility, we benchmark three representative tasks: cross-domain recommendation, highlight prediction, and questionnaire-based recommendation.\footnote{Our code is available at: https://github.com/imgkkk574/KuaiLive-M3}

\subsection{Cross-Domain Recommendation}

\subsubsection{Experimental Settings} 

Following previous studies~\cite{qu2025bridging, rappaz2021recommendation-liverec, cao2022disencdr}, we apply 5-core filtering separately to the live streaming and short video domains during data preprocessing. The filtered data contain 7,075,864 interactions among 21,762 users and 302,643 authors in the live streaming domain, as well as 49,881,858 interactions among 21,823 users and 2,686,313 videos in the short video domain.
To construct positive samples, we retain valid play records in the live streaming domain and interactions with a watch progress exceeding 10\% in the short video domain. We chronologically split the interactions into training, validation, and test sets at a ratio of 8:1:1. In the live streaming domain, authors are treated as target items for recommendation. We evaluate all methods using Recall@$K$ and NDCG@$K$, where $K \in \{10, 20, 40\}$.

\subsubsection{Benchmarked Methods} 

To provide a fair and comprehensive benchmark, we evaluate both representative single-domain recommendation methods and CDR methods. 
The single-domain methods include BPRMF~\cite{rendle2012bpr}, NeuMF~\cite{he2017neural}, NGCF~\cite{wang2019neural}, and LightGCN~\cite{he2020lightgcn}. The CDR methods include CMF~\cite{CMF}, CLFM~\cite{gao2013cross}, EMCDR~\cite{EMCDR}, CoNet~\cite{CoNet}, DCDCSR~\cite{zhu2018deep-DCDCSR}, DeepAPF~\cite{yan2019deepapf}, DTCDR~\cite{DTCDR}, BiTGCF~\cite{BiTGCF}, and MGCCDR~\cite{qu2025bridging}. 
Among these methods, MGCCDR is the only one specifically designed for live streaming recommendation. 
It facilitates cross-domain knowledge transfer by leveraging both overlapping users and non-overlapping items.

\subsubsection{Implementation Details} 
We implement the baseline methods using RecBole~\cite{zhao2021recbole, zhao2022recbole} and the official implementations released by their authors. 
For a fair comparison, we set the embedding dimension to 64 and the batch size to 2,048 for all methods.
We use the AdamW optimizer for training, with the learning rate searched over \{1e-3, 5e-3, 1e-4, 5e-4, 1e-5\}, the weight decay over \{1e-3, 1e-4, 1e-5\}, and the number of layers over \{1, 2, 3\}.
For each method, we select the hyperparameter configuration that achieves the best performance on the validation set.

\subsubsection{Experimental Results and Analysis} 
Table~\ref{tab:cross-domain} reports the overall performance of the evaluated single-domain and CDR methods on KuaiLive-M3. We make the following observations.
First, MGCCDR achieves the best overall performance. This result demonstrates the effectiveness of a method specifically designed for live streaming recommendation, which explicitly models the heterogeneous relations among users, streamers, and short videos to facilitate cross-domain knowledge transfer.
Second, LightGCN achieves the second-best performance and substantially outperforms all CDR methods except MGCCDR. One possible explanation is that KuaiLive-M3 contains tens of millions of interactions collected over a 4-week period, providing abundant collaborative signals for LightGCN to learn effective user and author representations through graph propagation.
These results also reveal the challenges of CDR on KuaiLive-M3. The short video and live streaming domains involve complex heterogeneous relations among users, authors, videos, and live rooms, while their item spaces are largely non-overlapping. Effectively exploiting this cross-domain structure therefore requires specialized model designs.

\subsection{Highlight Prediction}

Following the previous study~\cite{deng2024multimodal-KLive}, we formulate highlight prediction as a next-segment scoring task. Specifically, given the segments observed up to the current time in a live room, the model uses only the available historical information to predict a continuous highlight score for the forthcoming segment, without accessing its content or subsequent feedback. 
This task enables recommender systems to anticipate potential highlight moments and proactively distribute live streaming content at appropriate times.

Formally, for segment $k$ in live room $r$, we define the live viewer retention rate (LVTR) as
\begin{equation}
    \mathrm{LVTR}_{r,k} = \frac{N^{\mathrm{stay}}_{r,k}}
    {\max\left(N^{\mathrm{entered}}_{r,k},1\right)},
\end{equation}
where $N^{\mathrm{entered}}_{r,k}$ denotes the number of recorded viewers who have entered live room $r$ no later than the beginning of segment $k$, and $N^{\mathrm{stay}}_{r,k}$ denotes the number of these viewers who remain in the live room until the end of the segment.

To measure active engagement, we further define the engagement density as
\begin{equation}
    \mathrm{ED}_{r,k} = \frac{N^{\mathrm{like}}_{r,k}+ 2N^{\mathrm{comment}}_{r,k}}{\max\left(N^{\mathrm{entered}}_{r,k},1\right)},
\end{equation}
where $N^{\mathrm{like}}_{r,k}$ and $N^{\mathrm{comment}}_{r,k}$ denote the numbers of like and comment events occurring during segment $k$, respectively. Comments are assigned a larger weight because they generally indicate stronger user engagement than likes.

We independently apply min--max normalization to LVTR and ED within each live room. 
Following~\cite{deng2024multimodal-KLive} and empirical tuning, the continuous supervision score of segment $k$ is then defined as
\begin{equation}
    y_{r,k}= 0.6\widetilde{\mathrm{LVTR}}_{r,k} + 0.4\widetilde{\mathrm{ED}}_{r,k}.
\end{equation}
This score jointly captures passive viewer retention and active user engagement. Given the information observed up to segment $k$, the highlight prediction model is trained to predict the score $y_{r,k+1}$ of the next segment. 
For evaluation metrics that require binary labels, segments with scores at or above the 70th percentile within the corresponding live room are labeled as highlights.

\subsubsection{Experimental Settings}
We select the 10,000 live rooms with the largest numbers of playing interactions and retain rooms containing at least five valid segments and at least five viewers. 
For each room, we retrieve its temporally ordered 128-dimensional segment-level multi-modal embeddings as model inputs. The timestamp associated with each embedding is treated as the segment end time; the segment start time is defined as the end time of the preceding segment, with the live-room start time used for the first segment. 
The corresponding playing, liking, and commenting logs are used exclusively to construct supervision labels.
The retained live rooms are ordered chronologically by their start timestamps and split at the room level into training, validation, and test sets at a ratio of 70\%, 10\%, and 20\%, respectively. 
We report room-level macro-averaged Kendall's $\tau$ at $\Delta\in\{0,0.2,0.4\}$~\cite{deng2024multimodal-KLive} to measure ranking consistency and mean Average Precision (mAP)~\cite{zhao2022antpivot} to evaluate binary highlight detection.

\begin{table}[t]
    \centering
    \caption{Performance comparison of highlight prediction methods on KuaiLive-M3. The best result is highlighted in bold, and the second-best result is underlined.}
    \label{tab:highlight}
    \small
    \begin{tabular}{lcccc}
    \toprule
    \multirow{2}{*}{\textbf{Methods}}
      & \multicolumn{3}{c}{\textbf{Kendall's $\tau$ $\uparrow$}}
      & \multirow{2}{*}{\textbf{mAP $\uparrow$}} \\
    \cmidrule(lr){2-4}
      & $\Delta=0$ & $\Delta=0.2$ & $\Delta=0.4$ & \\
    \midrule
    MLP
      & 0.1474 & 0.1248 & 0.0872 & 0.5149 \\
    GRU
      & \textbf{0.5068} & \textbf{0.3070} & 0.2777 & \textbf{0.7776} \\
    AntPivot-style
      & \underline{0.5000} & \underline{0.3055} & \textbf{0.2917} & 0.7744 \\
    KuaiHL-style
      & 0.4988 & 0.3042 & \underline{0.2887} & \underline{0.7747} \\
    \bottomrule
    \end{tabular}
\end{table}

\subsubsection{Benchmarked Methods}
Research on highlight prediction for live-streaming content remains limited~\cite{deng2024multimodal-KLive, zhao2022antpivot}, and, to the best of our knowledge, the implementations of existing methods are not publicly available. 
We therefore implement two representative methods, KuaiHL~\cite{deng2024multimodal-KLive} and AntPivot~\cite{zhao2022antpivot}, based on their published architectures and adapt them to the next-segment prediction setting. 
We additionally include MLP and GRU as non-sequential and sequential baselines, respectively.
Specifically, MLP independently maps the current segment embedding to the highlight score of the next segment without modeling temporal dependencies.
GRU processes the observed segment sequence from left to right, allowing each next-segment prediction to depend on the current and preceding segments. 
The AntPivot-style hierarchical Transformer first employs causal local window-based attention to capture short-range transitions and then applies causal global self-attention to model room-level dependencies. 
The KuaiHL-style causal Transformer directly applies causal self-attention over the observed sequence.

\subsubsection{Implementation details}
For a fair comparison, all methods take exactly the same 128-dimensional segment embeddings as input and predict a highlight score for each segment.
We optimize all methods using AdamW with cosine learning-rate decay for at most 100 epochs. The batch size is set to 512. Early stopping is performed according to the validation mAP with a patience of 10 epochs. We search the learning rate over \{1e-3, 5e-3, 1e-4, 5e-4, 1e-5\}, and the weight decay over \{1e-5, 1e-4, 1e-3\}.
For each method, the hyperparameter configuration achieving the highest validation mAP is selected for test-set evaluation.

\subsubsection{Experimental Results and Analysis}

Table~\ref{tab:highlight} presents the performance of the benchmarked methods on the next-segment highlight prediction task. 
We make the following observations. 
First, all sequential models substantially outperform MLP, demonstrating that the current segment representation alone is insufficient and that temporal dependencies among previously observed segments are essential for anticipating future highlights. 
Second, GRU achieves the best mAP and Kendall's $\tau$ at $\Delta=0$ and $\Delta=0.2$, indicating that its recurrent structure effectively captures the evolving dynamics of live content. 
Third, the AntPivot-style model obtains the best Kendall's $\tau$ at the stricter threshold $\Delta=0.4$, suggesting that hierarchical temporal modeling is particularly effective at distinguishing segment pairs with large differences in highlight scores. 
Nevertheless, neither Transformer-based method consistently outperforms GRU, suggesting that modeling rapid and irregular temporal changes remains a key challenge for highlight prediction.

\subsection{Questionnaire-Based Recommendation}

\subsubsection{Experimental Settings}
Following previous work~\cite{zhang2024saqrec}, we construct the questionnaire-based recommendation benchmark exclusively from the live streaming domain, using play logs as implicit feedback and first-level questionnaire responses as explicit feedback, with streamers as target items.
Responses of ``Recommend whenever the streamer goes live'' and ``Recommend occasionally'' are treated as positive feedback, whereas ``Do not recommend this streamer'' is treated as negative.
We apply full-period 5-core filtering and chronologically split the interactions using a leave-one-out protocol. 
For evaluation, each held-out positive streamer is ranked against 99 randomly sampled streamers with which the user has never interacted, yielding a candidate set of 100 per test case~\cite{zhang2024saqrec}. 
Additional evaluation results using the full candidate pool of streamers are reported in Appendix~\ref{app:full candi eval}.
We report HitRatio@\{1, 5, 10\}, NDCG@\{5, 10\}, and MRR.

\subsubsection{Benchmarked Methods}
To ensure a comprehensive benchmark, we evaluate two categories of methods: single-behavior sequential recommendation methods and multi-behavior methods.
The single-behavior sequential methods include Caser~\cite{tang2018personalized-caser}, HGN~\cite{ma2019hierarchical-HGN}, NARM~\cite{li2017neural-NARM}, GRU4Rec~\cite{hidasi2015session}, SASRec~\cite{kang2018self}, and FMLP-Rec~\cite{zhou2022filter}. 
The multi-behavior methods include DMT~\cite{gu2020deep-DMT}, DFN~\cite{xie2021deep-DFN}, FeedRec~\cite{wu2022feedrec}, and SAQRec~\cite{zhang2024saqrec}. 
Among these, DMT is designed for general multi-behavior recommendation; DFN and FeedRec aim to integrate explicit and implicit user feedback; and SAQRec is specifically  designed for questionnaire-based recommendation. SAQRec first trains a propensity model and a satisfaction prediction model from questionnaire feedback, and then uses the resulting satisfaction model as a frozen teacher to provide satisfaction-aware supervision for the final recommender. 
Following previous work~\cite{zhang2024saqrec}, we further augment GRU4Rec, SASRec, and FMLP-Rec with questionnaire feedback as an additional behavior signal, yielding the variants GRU4Rec$_M$, SASRec$_M$, and FMLP-Rec$_M$, respectively.


\subsubsection{Implementation Details}
We implement all baseline methods using the official implementations released by their authors. We use an embedding dimension of 64 for all methods. Unless otherwise stated, the batch size is set to 4,096 on a single GPU.
We optimize all models using Adam. For the final recommendation model of each method, we search the learning rate over $\{1e-3, 5e-4, 1e-4\}$ and the weight decay over $\{0, 1e-6, 1e-5\}$. For models with configurable encoder depth, including FeedRec, SASRec, FMLP-Rec, SASRec$_M$, FMLP-Rec$_M$, DFN, and DMT, we additionally search the number of encoder blocks over $\{1,2,3\}$. 
We train for at most 100 epochs with an early-stopping patience of 10, and select the checkpoint and hyperparameter configuration with the best validation NDCG@10 for final test evaluation.

\begin{table}[t]
\centering
\caption{Performance comparison of sequential and questionnaire-aware methods for streamer recommendation on KuaiLive-M3. N@K and H@K denote NDCG@K and HitRatio@K, respectively. Bold and underlined values indicate the best and second-best results, respectively.}
\label{tab:questionnaire}
\footnotesize
\setlength{\tabcolsep}{4pt}
\begin{tabular}{llccccccc}
\toprule
\textbf{Cat.} & Model
  & N@5 & N@10 & H@1 & H@5 & H@10 & MRR \\
\midrule
\multirow{6}{*}{\textbf{Seq.}}
  & Caser    & 0.8982 & 0.9047 & 0.8348 & 0.9468 & 0.9667 & 0.8860 \\
  & HGN      & 0.8847 & 0.8924 & 0.8060 & 0.9457 & 0.9695 & 0.8687 \\
  & NARM     & 0.9037 & 0.9104 & 0.8475 & 0.9489 & 0.9695 & 0.8927\\
  & GRU4Rec  & 0.8942 & 0.9022 & 0.8326 & 0.9431 & 0.9673 & 0.8826 \\
  & SASRec   & 0.9153 & 0.9211 & 0.8646 & 0.9554 & 0.9731 & 0.9054 \\
  & FMLP-Rec & 0.9190 & 0.9248 & 0.8694 & 0.9583 & 0.9758 & 0.9093\\
\midrule
\multirow{7}{*}{\shortstack[l]{\textbf{Multi-}\\\textbf{beh.}}}
  & DMT          & 0.9117 & 0.9177 & 0.8678 & 0.9477 & 0.9662 & 0.9037 \\
  & DFN          & 0.9203 & 0.9255 & 0.8671 & \textbf{0.9618} & \textbf{0.9778} & 0.9095 \\
  & FeedRec      & 0.9120 & 0.9176 & 0.8574 & 0.9551 & 0.9723 & 0.9011 \\
  & GRU4Rec$_M$  & 0.9063 & 0.9126 & 0.8511 & 0.9497 & 0.9689 & 0.8957 \\
  & SASRec$_M$   & \textbf{0.9231} & \textbf{0.9281} & \textbf{0.8742} & \underline{0.9613} & \underline{0.9768} & \textbf{0.9133} \\
  & FMLP-Rec$_M$ & \underline{0.9216} & \underline{0.9269} & \underline{0.8722} & 0.9602 & 0.9762 & \underline{0.9119} \\
  & SAQRec       & 0.9091 & 0.9150 & 0.8536 & 0.9533 & 0.9714 & 0.8979 \\
\bottomrule
\end{tabular}
\vspace{-0.4cm}
\end{table}


\subsubsection{Experimental Results and Analysis}\label{sec:table6}
Table~\ref{tab:questionnaire} compares single-behavior sequential recommenders with questionnaire-aware multi-behavior methods. We make the following observations.
First, questionnaire feedback consistently improves the corresponding sequential backbones. 
Compared with their implicit-only counterparts, GRU4Rec$_M$, SASRec$_M$, and FMLP-Rec$_M$ achieve average relative improvements of 1.18\%, 0.77\%, and 0.23\%, respectively, across all six reported ranking metrics. 
In particular, SASRec$_M$ achieves the best overall performance on NDCG@5, NDCG@10, HR@1, and MRR, while DFN obtains the best HR@5 and HR@10. 
This consistently positive improvement demonstrates that even sparse questionnaire feedback provides complementary preference signals beyond implicit play behaviors.
Second, FMLP-Rec is the strongest implicit-only sequential baseline, indicating the effectiveness of frequency-domain sequential modeling. Nevertheless, its questionnaire-enhanced variant, FMLP-Rec$_M$, further improves all reported metrics, confirming the benefit of incorporating explicit satisfaction signals.
Finally, SAQRec does not outperform the strongest implicit-only or questionnaire-aware baselines in this setting. In particular, its NDCG@10 is lower than that of FMLP-Rec and SASRec$_M$. 
We attribute this to the inherent sparsity of questionnaire supervision relative to implicit behavioral signals. 
The dataset contains 35.8M implicit play logs in the live streaming domain, yet only 25,403 questionnaire responses, and this severe imbalance means that the propensity model and satisfaction predictor in SAQRec receive insufficient explicit training signal to generalize across the full streamer candidate pool, making the teacher-based alignment mechanism prone to introducing noise rather than reliable preference signals.
This result reveals an important open challenge for questionnaire-based recommendation: how to effectively exploit high-quality but extremely sparse explicit feedback without degrading performance on the uncovered majority.

\section{Potential Research Directions}

Our benchmark results demonstrate that KuaiLive-M3 effectively supports the three benchmarked tasks while revealing unique challenges inherent to live streaming recommendation. 
Beyond these tasks, its diverse logs, rich multi-modal content, and questionnaire feedback enable a broader range of research directions:

\paragraph{\textbf{Staytime Prediction.}}
For live streaming recommendation, accurate staytime prediction is critical for timely promoting relevant live rooms to target users and serves as a key user engagement metric in industrial systems~\cite{zhao2025towards, lu2025liveforesighter, liang2024ensure, lwhdiscovering}. 
Our KuaiLive-M3 provides fine-grained interaction timestamps and segment-level content embeddings, enabling further research and evaluation of staytime prediction models under realistic multi-stream settings.

\paragraph{\textbf{Generative Recommendation for Live Streaming.}}
Generative recommendation models quantize item content embeddings into discrete semantic IDs and employ generative models to predict the semantic ID of the next item~\cite{rajput2023recommender, wang2026onelive, shi2026ssrlive, deng2025onerec}.
Generative recommendation research in the live streaming domain remains limited~\cite{wang2026onelive, shi2026ssrlive}. The multi-modal content embeddings provided by KuaiLive-M3 can serve as a foundation for semantic ID construction, enabling the application of generative recommendation models to this domain. 
Notably, the segment-level embeddings further open avenues for generative recommendation in dynamic environments, where the evolving content of a live stream over time must be taken into account~\cite{guo2026room, shi2026ssrlive}. 
This creates opportunities for novel directions such as dynamic semantic ID modeling.

\paragraph{\textbf{Large Language Model for Recommendation.}}
Owing to their broad world knowledge and remarkable in-context learning and reasoning capabilities, large language models (LLMs) have been increasingly applied to recommendation tasks~\cite{uncovering, zhang2024large, team2026onereason, liu2025onerec-think, hou2026bridging, xiao2026scaling}. 
KuaiLive-M3 dataset opens up multiple possibilities for studying LLMs for recommendation in live streaming scenarios. The detailed metadata provided by KuaiLive-M3 facilitates the transfer of LLMs' world knowledge and reasoning capabilities to live streaming recommendation. Its cross-domain setting further supports research on LLM-enhanced cross-domain recommendation. 
Moreover, the two-level explicit questionnaire design enables the use of LLMs to simulate and generalize user preference signals, supporting deeper exploration of users' genuine interests. 

\section{Limitations}
While KuaiLive-M3 provides a rich testbed for live streaming recommendation research, several limitations should be noted. 
First, the questionnaire is distributed uniformly at random during normal platform usage, so respondents may differ systematically from non-respondents in engagement level or content preference, introducing potential selection bias in the explicit feedback signals.
Second, interactions in the short video domain are randomly downsampled for scalability, which may reduce behavioral signal density for certain users and widen the gap between the benchmark setting and real-world cross-domain scenarios.
Finally, the multimodal embeddings are extracted from an industrial MLLM that is not publicly available, which may limit reproducibility for embedding-sensitive tasks.
Despite these limitations, we believe KuaiLive-M3 constitutes a valuable and realistic resource for advancing live streaming recommendation research.

\section{Conclusion}
In this work, we present KuaiLive-M3, a multi-modal, multi-domain, and multi-feedback dataset for live streaming recommendation collected from Kuaishou. 
Unlike existing live streaming datasets, KuaiLive-M3 provides interaction logs and multi-modal content embeddings in both the short video and live streaming domains, and further incorporates explicit user feedback through a streamer recommendation frequency preference questionnaire administered to users in the live streaming domain. 
These properties enable research on live streaming recommendation tasks that more closely reflect industrial scenarios, including cross-domain recommendation, highlight prediction, and questionnaire-based recommendation. 
We provide a detailed description of the dataset construction process, key statistics, interaction patterns, and cross-domain transfer characteristics, and conduct benchmark experiments on the three aforementioned tasks. 
The experimental results offer insights and baselines for future research. 
We believe that KuaiLive-M3 will serve as a valuable resource and contribute to the advancement of live streaming recommendation research.
\label{sec:conclusion}

\begin{acks}
This work was supported by Beijing Natural Science Foundation (No. L2602179) and partially supported by the National Natural Science Foundation of China (No. 62376275, 62472426). Work partially done at Beijing Key Laboratory of Research on Large Models and Intelligent Governance, and Engineering Research Center of Next-Generation Intelligent Search and Recommendation, MOE. Supported by fund for building world-class universities (disciplines) of Renmin University of China.
Supported by Kuaishou Technology.
\end{acks}

\bibliographystyle{ACM-Reference-Format}
\bibliography{bib/references}
\appendix

\section{Evaluation on the Full Candidate Pool}
\label{app:full candi eval}
\begin{table}[t]
    \centering
    \caption{Performance comparison of sequential and questionnaire-aware methods for streamer recommendation on KuaiLive-M3 under full-candidate-pool evaluation.  Bold and underlined values indicate the best and second-best results, respectively.}
    \label{tab:questionnaire_full}
    \footnotesize \setlength{\tabcolsep}{4pt}
    \begin{tabular}{llcccccc}
    \toprule
    \textbf{Cat.} & Model
      & N@5 & N@10 & H@1 & H@5 & H@10 & MRR \\
    \midrule
    \multirow{5}{*}{\textbf{Seq.}}
      & HGN      & 0.2095 & 0.2244 & 0.1604 & 0.2545 & 0.3008 & 0.2085 \\
      & NARM     & 0.2935 & 0.3094 & 0.2372 & 0.3438 & 0.3930 & 0.2912 \\
      & GRU4Rec  & 0.1918 & 0.2108 & 0.1310 & 0.2474 & 0.3060 & 0.1907 \\
      & SASRec   & 0.2461 & 0.2641 & 0.1887 & 0.2976 & 0.3536 & 0.2453 \\
      & FMLP-Rec & 0.2616 & 0.2787 & 0.2039 & 0.3141 & 0.3668 & 0.2602 \\
    \midrule
    \multirow{4}{*}{\shortstack[l]{\textbf{Multi-}\\\textbf{beh.}}}
      & GRU4Rec$_M$  & 0.2150 & 0.2354 & 0.1498 & 0.2744 & 0.3376 & 0.2133 \\
      & SASRec$_M$   & \underline{0.3583} & \underline{0.3741} & \underline{0.2995} & \underline{0.4112} & \underline{0.4598} & \underline{0.3549} \\
      & FMLP-Rec$_M$ & \textbf{0.3630} & \textbf{0.3788} & \textbf{0.3069} & \textbf{0.4130} & \textbf{0.4619} & \textbf{0.3602} \\
      & SAQRec       & 0.1976 & 0.2197 & 0.1296 & 0.2606 & 0.3292 & 0.1963 \\
    \bottomrule
    \end{tabular}
    \vspace{-0.2cm}
\end{table}

\begin{table}[t]
\centering
\caption{Impact of short-video information on MGCCDR. Relative changes are computed with respect to the variant without short-video information. Bold values indicate the better result for each metric.}
\label{tab:cross-domain-ablation}
\footnotesize
\setlength{\tabcolsep}{3pt}
    \begin{tabular}{@{}lcccccc@{}}
    \toprule
    Model & R@10 & N@10 & R@20 & N@20 & R@40 & N@40 \\
    \midrule
    MGCCDR w/o video
    & 0.0731 & 0.1591 & 0.1153 & 0.1520
    & \textbf{0.1669} & 0.1570 \\
    MGCCDR
    & \textbf{0.0817} & \textbf{0.1795}
    & \textbf{0.1164} & \textbf{0.1651}
    & 0.1579 & \textbf{0.1633} \\
    \midrule
    Relative change
    & $+11.76\%$ & $+12.82\%$ & $+0.95\%$
    & $+8.62\%$ & $-5.39\%$ & $+4.01\%$ \\
    \bottomrule
    \end{tabular}
    \vspace{-0.3cm}
\end{table}

As shown in the table, evaluation on the full candidate pool reveals more pronounced performance differences between models, while the main findings remain largely consistent with those in Section~\ref{sec:table6}.
First, incorporating questionnaire feedback consistently improves the corresponding sequential backbones. GRU4Rec$_M$, SASRec$_M$, and FMLP-Rec$_M$ achieve average relative improvements of 11.87\%, 43.14\%, and 36.84\%, respectively, over their corresponding backbones across the six ranking metrics. These results suggest that questionnaire feedback can provide valuable explicit signals for recommendation despite its sparsity.
Second, FMLP-Rec$_M$ achieves the strongest overall performance among the evaluated models.
Finally, SAQRec underperforms the strongest implicit-only and questionnaire-aware baselines, ranking among the weakest methods in this setting. This suggests that its dedicated design may struggle to exploit the sparse questionnaire feedback available in KuaiLive-M3. Effectively using high-quality but extremely sparse explicit feedback therefore remains an important research challenge, for which KuaiLive-M3 provides a useful testbed.

\section{Impact of Short-Video Information on Cross-Domain Recommendation}
\label{app:cross-domain-ablation}

The cross-domain recommendation benchmark treats streamers as recommendation targets and evaluates whether users' short-video behaviors can improve the modeling of their preferences in the live-streaming domain. To quantify the contribution of short-video information, we compare MGCCDR with an ablated variant that excludes this information.

As shown in Table~\ref{tab:cross-domain-ablation}, incorporating short-video information improves five of the six reported metrics.
The largest relative improvements occur at the smallest evaluated cutoff, with gains of 11.76\% in Recall@10 and 12.82\% in NDCG@10.
However, Recall@40 decreases by 5.39\%, indicating that the benefits are not uniform across metrics and cutoffs. These results suggest that short-video information can improve live-streaming recommendation in the evaluated setting, particularly near the top of the ranked list.

\begin{table*}[t]
\centering
\caption{Overview of the shared and live streaming tables in KuaiLive-M3.}
\label{tab:dataset_fields_live}
\begin{threeparttable}
\footnotesize
\setlength{\tabcolsep}{5pt}
\renewcommand{\arraystretch}{1.15}
\begin{tabularx}{\textwidth}{p{0.20\textwidth}p{0.31\textwidth}Y}
\toprule
\textbf{Table} & \textbf{Description} & \textbf{Fields} \\
\midrule

\texttt{author\_profile.csv}
&
Profile attributes of creators appearing in either the live streaming
or short video domain.
&
\texttt{author\_id}, \texttt{is\_photo\_author},
\texttt{is\_live\_author}, \texttt{gender}, \texttt{age\_segment},
\texttt{fans\_user\_num}
\\

\texttt{user\_id\_set.csv}
&
The anonymized user population shared across the two domains.
&
\texttt{user\_id}
\\

\midrule
\multicolumn{3}{l}{\textit{live streaming domain}} \\
\midrule

\texttt{live\_interaction.csv}
&
Aggregated user viewing sessions, including entry context, watch
duration, and multiple interaction behaviors.
&
\texttt{user\_id}, \texttt{author\_id}, \texttt{live\_id},
\texttt{live\_play\_start\_timestamp},
\texttt{live\_play\_end\_timestamp}, \texttt{p\_date},
\texttt{live\_source\_category}, \texttt{enter\_live\_action},
\texttt{is\_auto\_play}, \texttt{is\_follow\_enter},
\texttt{is\_follow\_leave}, \texttt{play\_duration},
\texttt{like\_cnt}, \texttt{comment\_cnt},
\texttt{send\_gift\_cnt}, \texttt{send\_gift\_num},
\texttt{follow\_author\_cnt},
\texttt{cancel\_follow\_author\_cnt}, \texttt{share\_cnt},
\texttt{report\_live\_cnt}
\\

\texttt{live\_show.parquet}
&
Live-room impressions that were shown to users without a subsequent
room entry, supporting exposure-aware evaluation.
&
\texttt{live\_id}, \texttt{user\_id}, \texttt{author\_id},
\texttt{show\_timestamp}
\\

\texttt{live\_comment.csv}
&
Timestamped textual comments posted by users in live rooms.
&
\texttt{live\_stream\_id}, \texttt{user\_id}, \texttt{author\_id},
\texttt{content}, \texttt{comment\_ts}
\\

\texttt{live\_like.csv}
&
Event-level like actions with millisecond timestamps.
&
\texttt{user\_id}, \texttt{live\_id}, \texttt{like\_timestamp}
\\

\texttt{live\_share.csv}
&
Event-level sharing actions performed in live rooms.
&
\texttt{user\_id}, \texttt{live\_id}, \texttt{author\_id},
\texttt{share\_timestamp}, \texttt{is\_share\_success}
\\

\texttt{live\_questionnaire.csv}
&
Questionnaire-based explicit feedback collected from users while
watching live streams.
&
\texttt{author\_id}, \texttt{live\_stream\_id}, \texttt{user\_id},
\texttt{select\_option}, \texttt{second\_select\_option}
\\

\texttt{live\_room\_set.csv}
&
The universe of anonymized live-room IDs.
&
\texttt{live\_id}
\\

\texttt{live\_room\_meta.parquet}
&
Live-room titles, streamer mappings, lifecycle timestamps, and
broadcast durations.
&
\texttt{live\_id}, \texttt{author\_id}, \texttt{live\_name},
\texttt{start\_timestamp}, \texttt{end\_timestamp},
\texttt{live\_duration}
\\

\texttt{live\_emb\_64.parquet}
&
Room-level multimodal content representations aggregated over complete
live streams.
&
\texttt{live\_id}, \texttt{embedding}
\\

\texttt{live\_emb\_128\_ts/}
&
Timestamped segment-level multimodal representations of temporally
evolving live-room content.
&
\texttt{live\_stream\_id}, \texttt{segment\_id},
\texttt{feature\_128}, \texttt{timestamp}
\\

\bottomrule
\end{tabularx}

\begin{tablenotes}
\footnotesize
\item The segment-level embeddings are stored in 18 Parquet shards.
The \texttt{timestamp} field denotes the end of a variable-length
segment, and only segments with at least one viewer are retained.
\end{tablenotes}
\end{threeparttable}
\end{table*}

\begin{table*}[t]
\centering
\caption{Overview of the short video tables in KuaiLive-M3.}
\label{tab:dataset_fields_photo}
\begin{threeparttable}
\footnotesize
\setlength{\tabcolsep}{5pt}
\renewcommand{\arraystretch}{1.15}
\begin{tabularx}{\textwidth}{p{0.20\textwidth}p{0.31\textwidth}Y}
\toprule
\textbf{Table} & \textbf{Description} & \textbf{Fields} \\
\midrule

\texttt{photo\_interaction.csv}
&
Daily aggregated user--video interactions, including playback,
engagement, following, and sharing behaviors.
&
\texttt{user\_id}, \texttt{photo\_id}, \texttt{author\_id},
\texttt{show\_cnt}, \texttt{complete\_play\_cnt},
\texttt{play\_progress}, \texttt{like\_cnt},
\texttt{cancel\_like\_cnt}, \texttt{direct\_comment\_cnt},
\texttt{reply\_comment\_cnt}, \texttt{comment\_stay\_duration},
\texttt{follow\_cnt}, \texttt{cancel\_follow\_cnt},
\texttt{share\_cnt}
\\

\texttt{photo\_play.parquet}
&
Event-level short video playback records with entry and exit
timestamps, playback states, and user--creator relationship states.
&
\texttt{user\_id}, \texttt{photo\_id}, \texttt{author\_id},
\texttt{enter\_timestamp}, \texttt{leave\_timestamp},
\texttt{enter\_play\_type}, \texttt{is\_complete\_play},
\texttt{like\_status\_type}, \texttt{is\_follow\_before\_play},
\texttt{is\_follow\_after\_play}
\\

\texttt{photo\_id\_set.csv}
&
The universe of anonymized short video IDs.
&
\texttt{photo\_id}
\\

\texttt{photo\_meta.parquet}
&
short video duration, creator mapping, display format, cover title,
OCR text, and subtitles.
&
\texttt{photo\_id}, \texttt{author\_id}, \texttt{duration},
\texttt{display\_type}, \texttt{cover\_title},
\texttt{video\_texts}
\\

\texttt{photo\_tag.csv}
&
A four-level hierarchical content taxonomy for short videos.
&
\texttt{photo\_id}, \texttt{first\_level\_category\_name},
\texttt{second\_level\_category\_name},
\texttt{third\_level\_category\_name},
\texttt{fourth\_level\_category\_name}
\\

\texttt{photo\_emb\_128.parquet}
&
Video-level multimodal content representations for short videos.
&
\texttt{photo\_id}, \texttt{feature}
\\

\bottomrule
\end{tabularx}

\begin{tablenotes}
\footnotesize
\item In \texttt{photo\_play.parquet}, a user--video pair may occur
multiple times because each row corresponds to an individual playback
event.
\end{tablenotes}
\end{threeparttable}
\end{table*}

\section{Dataset Schema and Field Descriptions}
\label{app:dataset_schema}

KuaiLive-M3 consists of 18 tables and collections covering shared entities, live streaming activities, and short video activities. Tables~\ref{tab:dataset_fields_live} and~\ref{tab:dataset_fields_photo} summarize the role and fields of each released data table. The shared tables define the user and creator populations across the two domains. 
The live streaming tables provide room metadata, exposure and interaction logs, questionnaire-based feedback, and room- and segment-level multi-modal representations. 
The short video tables contain metadata, aggregated interactions, event-level playback records, hierarchical content categories, and video-level multimodal representations.

Identifiers are anonymized consistently across related tables, allowing users, authors, live rooms, and short videos to be joined through their corresponding ID fields.
All timestamps are provided with millisecond precision unless otherwise specified. 
For the segment-level live streaming representations, each live stream is partitioned into equal-duration segments, although the segment duration may vary across live rooms. We retain only segments associated with at least one user interaction, so the released segments may be non-consecutive within a live room. The stored timestamp denotes the end of a variable-length segment, enabling the embeddings to be aligned with temporally overlapping user interactions.
Detailed field definitions, data types, and preprocessing notes are available on the project website.\footnote{\url{https://imgkkk574.github.io/KuaiLive-M3/description/}}

\end{document}